\documentclass[sigconf]{acmart}

\usepackage{multirow}
\usepackage{booktabs}
\usepackage{algorithmic}
\usepackage{algorithm}
\usepackage{array}
\usepackage{booktabs}
\usepackage{multirow}
\usepackage{amsmath}
\usepackage{subfig}
\usepackage{wasysym}
\usepackage{caption}
\usepackage{multirow}
\usepackage{algorithmic}

\usepackage{color}

\usepackage{subcaption}

\AtBeginDocument{%
  }

\setcopyright{acmlicensed}
\copyrightyear{2018}
\acmYear{2018}
\acmDOI{XXXXXXX.XXXXXXX}
\acmConference[Conference acronym 'XX]{Make sure to enter the correct
  conference title from your rights confirmation email}{June 03--05,
  2018}{Woodstock, NY}
\acmISBN{978-1-4503-XXXX-X/2018/06}

\usepackage{fancyhdr}
\fancypagestyle{empty}{%
  \fancyhf{} 
  \fancyhead[L]{}
  \fancyhead[C]{}
  \fancyhead[R]{}
  \fancyfoot[L]{}
  \fancyfoot[C]{}
  \fancyfoot[R]{}
}
\fancypagestyle{empty}

\begin{document}

\title{SCI: A Simple and Effective Framework for Symmetric Consistent Indexing in Large-Scale Dense Retrieval}

\author{Huimu Wang\textsuperscript{\rm 1}, Yiming Qiu\textsuperscript{\rm 1,$\dagger$}, Xingzhi Yao\textsuperscript{\rm 1, $\dagger$}, Zhiguo Chen\textsuperscript{\rm 1}, \\  Guoyu Tang\textsuperscript{\rm 1}, Songlin Wang\textsuperscript{\rm 1}, Sulong Xu\textsuperscript{\rm 1},  Mingming Li\textsuperscript{\rm 2}\\}
\thanks{$\dagger$ Corresponding author.}
\affiliation{
  \institution{\textsuperscript{\rm 1}JD.com \country{China},\textsuperscript{\rm 2} Institute of Information Engineering, Chinese Academy of Sciences \country{China} }
   }
\email{liemingming@outlook.com}
\email{{wanghuimu1, qiuyiming3, yaoxingzhi1,chenzhiguo, tangguoyu, wangsonglin3, xusulong}@jd.com}

\renewcommand{\shortauthors}{Wang et al.}

\begin{abstract}
  Dense retrieval has become the industry standard in large-scale information retrieval systems due to its high efficiency and competitive accuracy. Its core relies on a coarse-to-fine hierarchical architecture that enables rapid candidate selection and precise semantic matching, achieving millisecond-level response over billion-scale corpora. This capability makes it essential not only in traditional search and recommendation scenarios but also in the emerging paradigm of generative recommendation driven by large language models, where semantic IDs-themselves a form of coarse-to-fine representation-play a foundational role. However, the widely adopted dual-tower encoding architecture introduces inherent challenges, primarily representational space misalignment and retrieval index inconsistency, which degrade matching accuracy, retrieval stability, and performance on long-tail queries. These issues are further magnified in semantic ID generation, ultimately limiting the performance ceiling of downstream generative models.

To address these challenges, this paper proposes a simple and effective framework named SCI comprising two synergistic modules: a symmetric representation alignment module that employs an innovative input-swapping mechanism to unify the dual-tower representation space without adding parameters, and an consistent indexing with dual-tower synergy module that redesigns retrieval paths using a dual-view indexing strategy to maintain consistency from training to inference. The framework is systematic, lightweight, and engineering-friendly, requiring minimal overhead while fully supporting billion-scale deployment. We provide theoretical guarantees for our approach, with its effectiveness validated by results across public datasets and real-world e-commerce datasets.
\end{abstract}



\received{20 February 2007}
\received[revised]{12 March 2009}
\received[accepted]{5 June 2009}

\maketitle
\section{Introduction}

\begin{figure}[t!]
\centering
\includegraphics[scale=0.2]{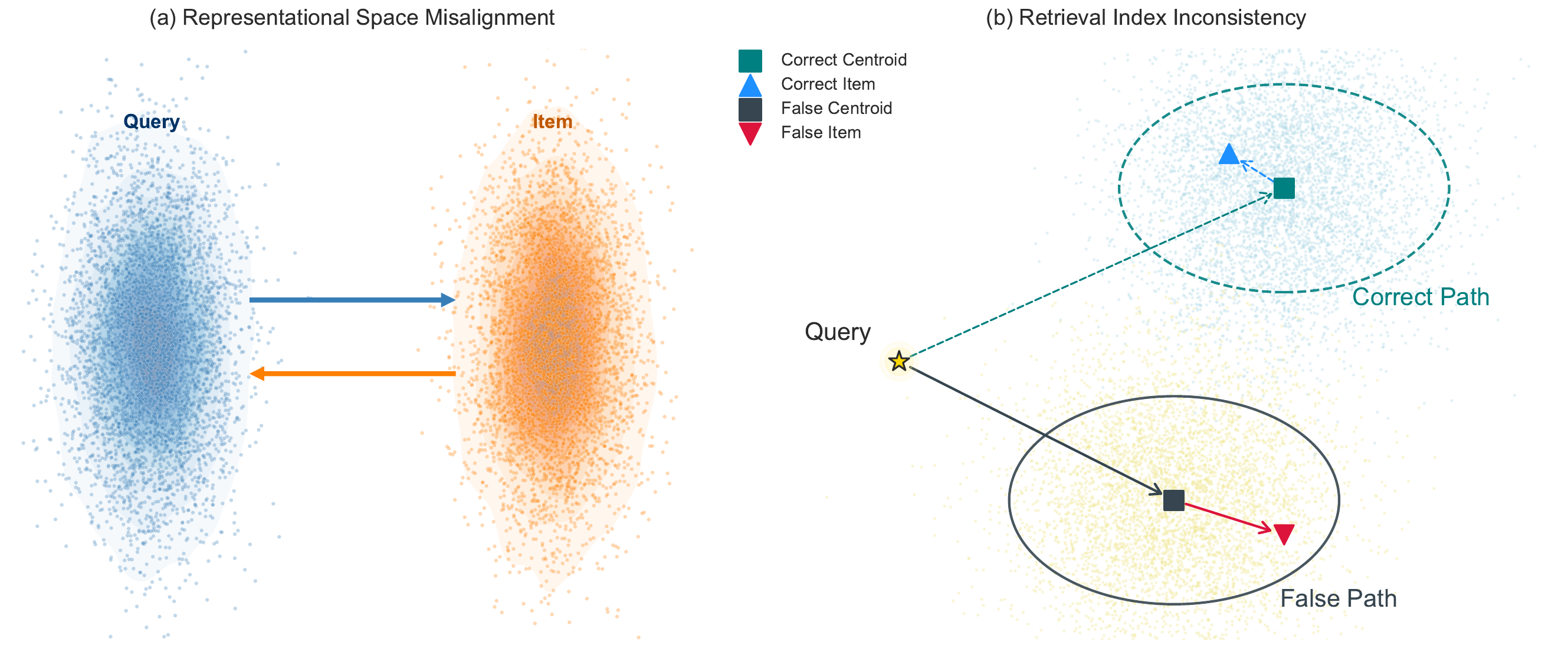}
\caption{The problem of space misalignment and index inconsistency in dual-tower retrieval.}
\label{fig:rep_Misalignment}
\end{figure}

In large-scale information retrieval/recommendation systems, dense retrieval has emerged as the industry standard due to its exceptional efficiency and competitive accuracy \cite{karpukhin2020dense, khattab2020colbert,lin2020distilling,xie2023t2ranking,dong2022incorporating,zhan2021optimizing}. At the heart of this approach lies a coarse-to-fine hierarchical retrieval architecture: by encoding queries and items into low-dimensional dense vectors, the system first rapidly narrows the search scope at a coarse level (e.g., via cluster centroids \cite{karpukhin2020dense}, graph navigation points \cite{malkov2018efficient}, or primary quantization of semantic IDs \cite{jegou2010product}), and subsequently performs precise semantic matching at a fine-grained level (e.g., through residual quantization or nearest-neighbor distance comparison \cite{xiong2020approximate}). This hierarchical strategy enables millisecond-level responses over billion-scale corpora, forming the foundation for its practical deployment \cite{chen2021spann}.

This capability not only makes dense retrieval highly effective in traditional recommendation and search scenarios but also establishes it as a cornerstone in the emerging paradigm of generative recommendation driven by large language models. It is particularly important to emphasize that semantic IDs—essential to generative models—are themselves a typical example of coarse-to-fine hierarchical representations. They are constructed using techniques such as hierarchical residual quantization (e.g., RQ-VAE \cite{rajput2023recommender, kuai2024breaking}). However, the quality of the resulting discrete vocabulary is entirely dependent on the semantic space provided by the upstream dense retrieval system. The effectiveness of representation learning in dense retrieval directly determines the coherence of semantic IDs across coarse partitioning and fine-grained quantization, thereby defining the performance ceiling for subsequent generative models \cite{kuai2024breaking}.

As the core architecture enabling dense retrieval, the dual-tower model learns separate vector representations for queries and items through independent encoder towers. This decoupled design is key to its efficiency and enables seamless integration with approximate nearest neighbor (ANN) search techniques \cite{xiong2020approximate}. However, this independent encoding mechanism also introduces inherent challenges, primarily manifested in two interrelated aspects:

1) Representational Space Misalignment: During training, the two towers operate with limited interactive supervision, causing their embeddings to occupy divergent vector subspaces and exhibit significant anisotropy (as shown in Fig \ref{fig:rep_Misalignment}). As a result, simple similarity measures often fail to capture true semantic relevance. Although contrastive learning and related techniques have been used to align positive pairs \cite{karpukhin2020dense,xiong2020approximate}, such alignment is often incomplete in practice or relies heavily on external data—as illustrated by the distribution shift between dual-tower embeddings.
2) Retrieval Index Inconsistency: The representational misalignment is further exacerbated during retrieval by the structure of the index. Popular indexing methods—including cluster-based IVF-PQ, graph-based HNSW, and semantic ID generation—all adopt a coarse-to-fine retrieval architecture. Due to space misalignment, the initial matching between a query and coarse-level structures (e.g., IVF centroids, HNSW entry points, or coarse semantic ID tokens) is biased from the start (query → irrelevant cluster center → items), confining subsequent fine-grained retrieval to incorrect neighborhoods and propagating errors throughout the process (as shown in Fig \ref{fig:rep_Misalignment}).

Critically, these inconsistencies become systematically embedded in the semantic ID generation process. Techniques like residual quantization \cite{barnes1996advances} learn a hierarchical discrete representation of items based solely on item-item relationships, independent of query distributions. When the dual-tower representations are misaligned, the resulting semantic ID vocabulary is built upon biased item representations. Consequently, generative models inherit a fundamentally flawed semantic space, severely limiting the system’s overall performance potential.

These inconsistencies impact system performance in three key areas: semantic matching accuracy suffers from misaligned spaces; retrieval stability is compromised by distributional drift; and long-tail queries perform poorly due to inadequate fine-grained semantic capture.

Therefore, resolving representational misalignment in dense retrieval is not only vital for improving traditional retrieval systems but also essential for ensuring high-quality semantic ID inputs in the era of generative recommendation systems. This paper delves into this core challenge and introduces a consistency-focused retrieval framework named SCI comprising two tightly integrated components:
1) \textbf{Symmetric Representation Alignment}: Introduces an innovative input-swapping mechanism—feeding query samples into the item tower to generate item-view query representations, and item samples into the query tower to obtain query-view item representations. By applying a symmetric contrastive loss, the dual towers naturally converge toward a unified semantic space while preserving their individual encoding capabilities. This approach requires only additional forward passes during training, introduces no new parameters, and significantly improves representation quality.
2) \textbf{Consistent Indexing with dual-tower Synergy (CI)}: Re-engineers retrieval path consistency. We propose a novel indexing strategy: when building a coarse-to-fine index (e.g., IVF-PQ), all item embeddings encoded by the query tower are used for coarse clustering, ensuring queries and centroids reside in the same semantic space. Then, within each cluster, the original item tower embeddings are used for residual quantization. This ensures the retrieval path remains aligned with the learning objective.

The proposed framework offers three key advantages:1) Systematic: Ensures end-to-end consistency from training to retrieval via coordinated design of representation learning and indexing. 2) Simple: Only introduces input swapping during training without complex external modules. 3) Engineering-Friendly: Minimal training overhead, and indexing is fully compatible with industrial ANN libraries, supporting billion-scale deployment.
To validate the framework, we performed comprehensive experiments on on two large publicly available
datasets and a real-world billion-scale e-Commerce data set. Offline evaluations demonstrated significant improvements in core metrics such as Recall and MAP. The main contributions of this work are summarized as follows:
\begin{itemize}
    \item We propose a novel consistency-oriented retrieval framework that systematically addresses alignment issues across the entire pipeline—from representation learning to retrieval execution.
    \item We introduce Symmetric Representation Alignment (SymmAligner), an efficient input-swapping mechanism for cross-tower alignment, and Consistent Indexing with dual-tower synergy (CI), a novel indexing architecture that incorporates query-side information into index construction.
    \item We provide extensive validations on large publicly and industrial datasets, offering a practical and scalable solution for optimizing large-scale dense retrieval systems.
\end{itemize}
\section{Related Works}

\subsection{Representation Learning}
The dual-tower model is the predominant architecture for large-scale dense retrieval due to its efficiency with pre-computable vectors for Approximate Nearest Neighbor (ANN) search \cite{yu2021dual,yang2020mixed,mao2021simplex,huang2013learning,su2023beyond}. However, its decoupled design inherently leads to a representation space misalignment between queries and items. Existing efforts to mitigate this, such as enhanced negative sampling \cite{karpukhin2020dense,xiong2020approximate}, late-interaction mechanisms \cite{su2023beyond,wang2025unleashing}, or knowledge distillation from cross-attention models \cite{chen2021salient}, often introduce significant scalability challenges, including increased inference latency or complex training procedures, limiting their practical deployment.

\subsection{Approximate Retrieval}
To achieve millisecond-level latency over billion-item corpora, ANN search \cite{johnson2019billion} is essential and universally relies on a coarse-to-fine hierarchical paradigm. This principle underpins clustering-based methods like IVF \cite{sivic2003video} and graph-based methods like HNSW \cite{malkov2018efficient}, which first narrow the search space to a promising subset before performing fine-grained comparisons. Modern data-driven approaches, such as Semantic IDs, also learn a hierarchical discrete partitioning of the semantic space, effectively internalizing this coarse-to-fine structure. The performance of this entire retrieval pipeline, however, is fundamentally dependent on the quality and consistency of the initial vector representations.

\begin{figure*}[htbp!]
\centering
\includegraphics[width=1.0\textwidth]{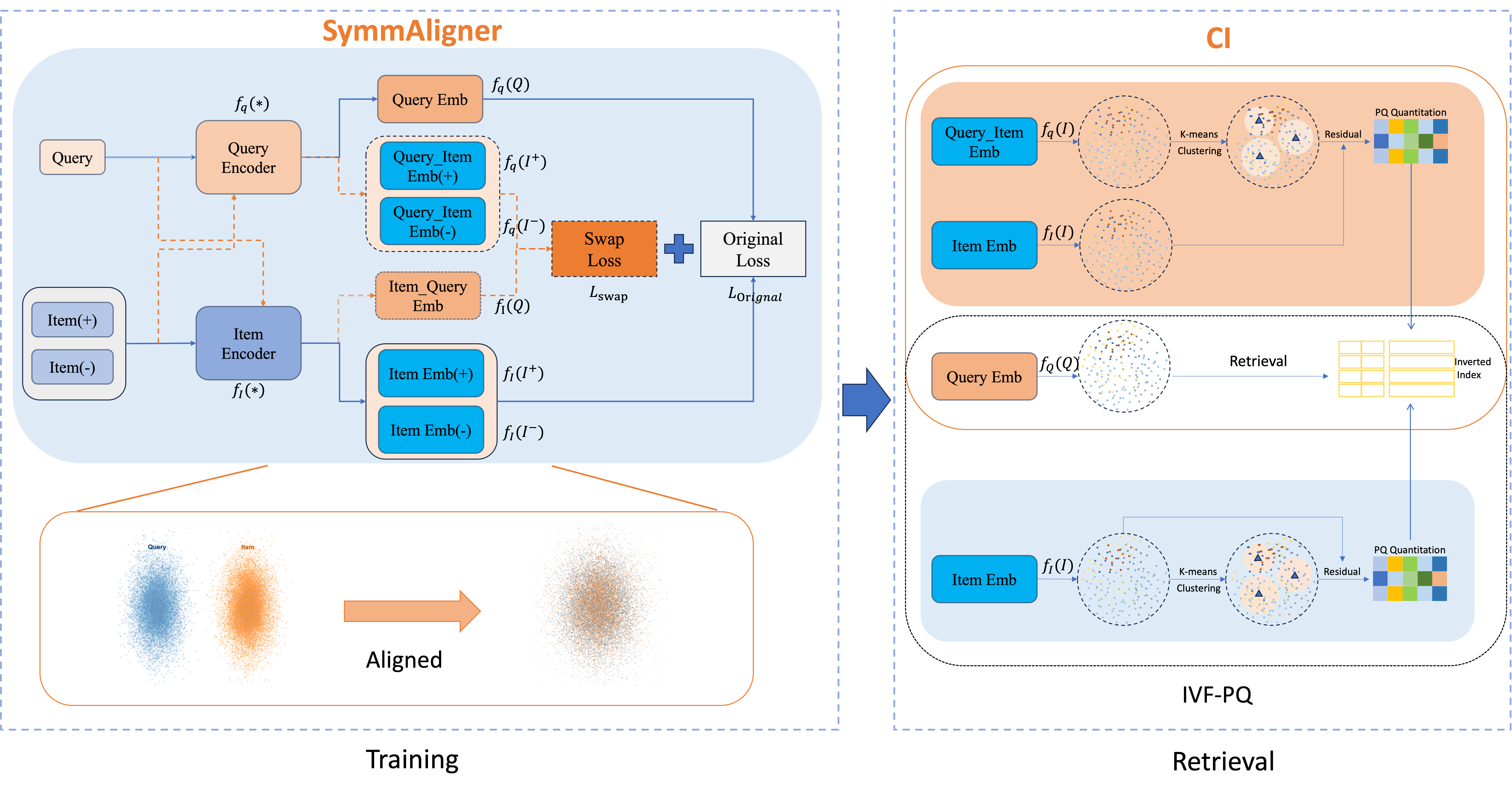}
\caption{The workflow of SCI: (a) \textbf{Symmetric Representation Alignment (SymmAligner)} via input swapping and joint optimization, (b) \textbf{Consistent Indexing with Dual-Tower Synergy (CI)} compared to traditional IVF-PQ, where CI uses the query encoder for unified representation with dual-tower enhancement.}
\label{fig:framework}
\end{figure*}

\subsection{Joint Index Optimization}
The optimization of representation learning and index retrieval has traditionally been decoupled. Some research has explored joint optimization, which integrates index structures or quantization losses directly into model training \cite{fang2022joint,zhan2021jointly,zhan2022learning,zhu2018learning}. However, these methods are often computationally intensive and exhibit poor compatibility with highly optimized, standard ANN libraries. Consequently, the industry standard remains a decoupled approach for its engineering simplicity and stability. The central challenge, which this paper addresses, is to resolve the foundational representation space misalignment within this practical, decoupled framework, ensuring consistency between the learned representations and the downstream retrieval index.

\section{Methodology}
\label{sec:method}

\subsection{Overview}

Our proposed framework, illustrated in Figure~\ref{fig:framework}, addresses dual-tower retrieval limitations through an integrated pipeline comprising two synergistic components. The complete workflow operates as follows: during training, the \textbf{Symmetric Representation Alignment (SymmAligner)} component performs input swapping between query and item encoders, computes an alignment loss, and optimizes model parameters to eliminate representation anisotropy. During indexing, the \textbf{Consistent Indexing with Dual-Tower Synergy (CI)} component processes all items through the query encoder to build cluster centers, then applies enhanced quantization using embeddings from both towers. For retrieval, queries are encoded through the query encoder and efficiently searched using the unified index.

The framework offers crucial advantages: it requires no architectural changes, introduces minimal computational overhead, maintains compatibility with various ANN algorithms, and demonstrates universal applicability across different encoder architectures (BERT, ResNet, etc.) and multimodal scenarios (text-image, text-video retrieval). The tight integration between symmetric training and consistent indexing ensures representation compatibility while preserving the efficiency benefits of dual-tower architectures.

\subsection{Symmetric Representation Alignment (SymmAligner)}

The fundamental challenge of dual-tower models lies in the incompatible representation spaces learned by the query encoder $f_q$ and item encoder $f_i$ during independent optimization. This incompatibility causes even semantically related queries and items to have distant embedding vectors in the representation space. To overcome this limitation, we propose a simple symmetric representation alignment method based on input swapping.

\subsubsection{Training Framework and Loss Formulation}

Our training framework builds upon the traditional dual-tower optimization pipeline while introducing a novel symmetric objective to enforce representation alignment. The standard approach relies on a triplet loss, which we refer to as the original loss ($\mathcal{L}_{\text{original}}$), to ensure task-specific performance. For a given training sample $(Q, I^+, I^-)$, this loss is formulated as:
\begin{equation}
\resizebox{.9\hsize}{!}{
$\mathcal{L}_{\text{original}}(\theta_q, \theta_i) = \mathbb{E}_{(Q,I^+,I^-)}\left[\max\left(0, \delta - S(f_q(Q), f_i(I^+)) + S(f_q(Q), f_i(I^-))\right)\right]$}
\end{equation}
While effective for the primary retrieval task, this objective does not explicitly prevent the query and item encoders from learning incompatible, anisotropic representations.

To overcome this limitation, we introduce an innovative input-swapping mechanism. For each training sample, in addition to the standard embeddings, we compute a set of swapped embeddings. We feed the query $Q$ into the item encoder to obtain an item-view query representation, denoted as $\mathbf{v}_{q,\text{swap}} = f_i(Q)$, and conversely, feed items $I^+, I^-$ into the query encoder to obtain query-view item representations, denoted as $\mathbf{v}_{i^+,\text{swap}} = f_q(I^+)$ and $\mathbf{v}_{i^-,\text{swap}} = f_q(I^-)$.

Based on these swapped representations, we construct a Swap Alignment Loss ($\mathcal{L}_{\text{swap}}$) that mirrors the original objective:
\begin{equation}
\resizebox{.9\hsize}{!}{
$\mathcal{L}_{\text{swap}}(\theta_q, \theta_i) = \mathbb{E}_{(Q,I^+,I^-)}\left[\max\left(0, \delta - S(\mathbf{v}_{q,\text{swap}}, \mathbf{v}_{i^+,\text{swap}}) + S(\mathbf{v}_{q,\text{swap}}, \mathbf{v}_{i^-,\text{swap}})\right)\right]$}
\end{equation}
This loss function forces the model to maintain correct similarity relationships even when the encoders' roles are reversed. This symmetric constraint incentivizes both encoders to learn a shared, compatible representation space that preserves semantic consistency regardless of the input's origin.

The complete training objective, our total loss ($\mathcal{L}_{\text{total}}$), combines both the standard retrieval objective and our novel alignment objective:
\begin{equation}
\mathcal{L}_{\text{total}}(\theta_q, \theta_i) = (1-\lambda)\mathcal{L}_{\text{original}}(\theta_q, \theta_i) + \lambda \mathcal{L}_{\text{swap}}(\theta_q, \theta_i)
\end{equation}
The hyperparameter $\lambda$ controls the balance of alignment strength, typically set in the range $[0.1, 1.0]$. This symmetric training approach offers several notable advantages: it preserves the model architecture with only additional forward computation overhead; requires no additional training data; works with various encoder types and multimodal scenarios; and enables flexible alignment control through a simple hyperparameter adjustment.

\subsubsection{Gradient Analysis and Theoretical Foundations}

Understanding how the swap loss promotes representation alignment by providing complementary gradient paths is crucial for deeply appreciating the effectiveness of our method. By analyzing the gradient propagation mechanism, we can clearly see how the swap loss provides additional optimization signals for the dual-tower model.

For query tower parameters $\theta_q$, the total gradient $\nabla_{\theta_q}\mathcal{L}_{\text{total}}$ synthesizes two distinct optimization paths:

\textbf{Original loss gradient path}:
\begin{equation}
\resizebox{.9\hsize}{!}{
$\nabla_{\theta_q}\mathcal{L}_{\text{original}} \propto \mathbb{E}\left[ \left( -\nabla_{\theta_q}S(f_q(Q), f_i(I^+)) + \nabla_{\theta_q}S(f_q(Q), f_i(I^-)) \right) \right]$}
\end{equation}

This path flows exclusively through the query tower's encoding of queries, training the query tower to become a specialized query encoder. It focuses on making query representations better interact with item representations, but doesn't directly address the problem of representation space alignment.

\textbf{Swap loss gradient path}:
\begin{equation}
\resizebox{.95\hsize}{!}{
$\nabla_{\theta_q}\mathcal{L}_{\text{swap}} \propto \mathbb{E}\left[ \left( -\nabla_{\theta_q}S(f_i(Q), f_q(I^+)) + \nabla_{\theta_q}S(f_i(Q), f_q(I^-)) \right) \right]$}
\end{equation}

This innovative path flows through the query tower's encoding of items, forcing the query tower to also learn meaningful representations for items. This symmetric effect similarly applies to the item tower, creating a bidirectional optimization mechanism that compels both towers to learn a shared universal semantic mapping.

\subsubsection{Theoretical Foundations}

The swap loss mechanism provides theoretical guarantees for representation alignment through complementary gradient paths and space regularization.

\begin{lemma}[Gradient Linear Independence]
\label{lemma:independence}
Under the condition that query and item distributions are not identical, the gradients $\nabla_{\theta_q}\mathcal{L}_{\text{original}}$ and $\nabla_{\theta_q}\mathcal{L}_{\text{swap}}$ are generally linearly independent. (Proof see Appendix \ref{app:lemma1})
\end{lemma}

\begin{lemma}[Representation Alignment] 
\label{lemma:alignment}
Minimizing $\mathcal{L}_{\text{total}}$ implicitly minimizes the alignment error (Proof see Appendix \ref{app:lemma2}):
\begin{equation}
\mathcal{A}(\theta_q, \theta_i) = \mathbb{E}_{Q,I}\left[ \left( S(f_q(Q), f_i(I)) - S(f_i(Q), f_q(I)) \right)^2 \right]
\end{equation}
\end{lemma} 

\begin{lemma}[Anisotropy Reduction]
\label{lemma:anisotropy} 
The swap loss regularizes the embedding space, reducing anisotropy by enforcing compatible covariance structures (Proof see Appendix \ref{app:lemma3}):
\begin{equation}
\text{Cov}(f_q(\mathcal{Q} \cup \mathcal{I})) \approx \text{Cov}(f_i(\mathcal{Q} \cup \mathcal{I}))
\end{equation}
\end{lemma}

Collectively, these lemmas establish that our symmetric training approach addresses the core limitations of dual-tower architectures. Lemma \ref{lemma:independence} ensures non-redundant optimization signals, Lemma \ref{lemma:alignment} guarantees representation space compatibility, and Lemma \ref{lemma:anisotropy} enables effective nearest neighbor search by creating an isotropic embedding space. These theoretical properties form the foundation for our consistent indexing framework.

\subsubsection{Linear Model Analysis}
To ground our theoretical results in concrete mathematical intuition, we analyze a simplified single-layer linear dual-tower model.

Consider linear transformations $f_q(Q) = W_q q$ and $f_i(I) = W_i i$. The total gradient reveals a symmetrization mechanism:
\begin{equation}
\frac{\partial \mathcal{L}_{\text{total}}}{\partial W_q} = W_i \left[ \Delta i q^\top + \lambda q \Delta i^\top \right]
\end{equation}
where $\Delta i = i^- - i^+$. This suggests the symmetrization operator:
\begin{equation}
M(q, \Delta i) = \frac{1}{2}(q \Delta i^\top + \Delta i q^\top)
\end{equation}

\textbf{Collapse analysis} reveals that the swap loss provides meaningful signals except in degenerate cases: (1) $\lambda = 0$ (no swap loss), (2) $q \parallel \Delta i$ (vector parallelism), or (3) special orthogonal conditions. In practice, these conditions are rarely satisfied since queries and items typically exhibit different statistical properties. (Detailed analysis see Appendix \ref{app:collapse})

\begin{algorithm}[t]
\caption{Symmetric Representation Alignment (SymmAligner) Training}
\label{alg:training}
\begin{algorithmic}[1]
\REQUIRE Query encoder $f_q$, item encoder $f_i$, dataset $\mathcal{D}$, margin $\delta$, alignment weight $\lambda$
\ENSURE Aligned encoders $f_q^*$, $f_i^*$
\STATE Initialize $f_q$, $f_i$ with standard pretrained weights
\FOR{each training batch $(Q, I^+, I^-)$}
    \STATE \textit{// Compute standard embeddings and loss}
    \STATE $\mathbf{v}_q \leftarrow f_q(Q)$; $\mathbf{v}_{i^+} \leftarrow f_i(I^+)$; $\mathbf{v}_{i^-} \leftarrow f_i(I^-)$
    \STATE $\mathcal{L}_{\text{original}} \leftarrow \max(0, \delta - S(\mathbf{v}_q, \mathbf{v}_{i^+}) + S(\mathbf{v}_q, \mathbf{v}_{i^-}))$
    
    \STATE \textit{// Compute swapped embeddings and loss}
    \STATE $\mathbf{v}_{q,\text{swap}} \leftarrow f_i(Q)$; $\mathbf{v}_{i^+,\text{swap}} \leftarrow f_q(I^+)$; $\mathbf{v}_{i^-,\text{swap}} \leftarrow f_q(I^-)$
    \STATE $\mathcal{L}_{\text{swap}} \leftarrow \max(0, \delta - S(\mathbf{v}_{q,\text{swap}}, \mathbf{v}_{i^+,\text{swap}}) + S(\mathbf{v}_{q,\text{swap}}, \mathbf{v}_{i^-,\text{swap}}))$
    
    \STATE \textit{// Combined optimization}
    \STATE $\mathcal{L}_{\text{total}} \leftarrow \mathcal{L}_{\text{original}} + \lambda \mathcal{L}_{\text{swap}}$
    \STATE Update $\theta_q, \theta_i$ using $\nabla\mathcal{L}_{\text{total}}$
\ENDFOR
\RETURN $f_q$, $f_i$
\end{algorithmic}
\end{algorithm}

\subsection{Consistent Indexing with Dual-Tower Synergy (CI)}

Building upon the representation alignment achieved through SymmAligner, we introduce the \textbf{Consistent Indexing with Dual-Tower Synergy (CI)} framework to eliminate the distribution mismatch in traditional dual-tower retrieval.

Traditional dual-tower retrieval suffers from inherent inconsistency: queries and items are encoded through different towers, yet compared during retrieval. This can be formalized as:
\begin{equation}
\{I_1, \ldots, I_k\} = \mathcal{A}\left(f_q(Q), \{f_i(I)\}_{I \in \mathcal{D}}\right)
\end{equation}
where the ANN algorithm $\mathcal{A}$ must bridge the gap between query space $f_q$ and item space $f_i$.

Our solution leverages the aligned representations from SymmAligner to create a unified indexing scheme:
\begin{equation}
\{I_1, \ldots, I_k\} = \mathcal{A}\left(f_q(Q); \underbrace{\{f_q(I)\}_{I \in \mathcal{D}}}_{\text{Coarse Structure}} , \underbrace{\{f_i(I)\}_{I \in \mathcal{D}}}_{\text{Fine Representation}}\right)
\end{equation}

The key innovation lies in assigning distinct roles to each tower: the query encoder defines the coarse index structure, while both towers contribute to fine-grained representations. This synergistic approach is enabled by the representation alignment (Lemma~\ref{lemma:alignment}) and isotropy (Lemma~\ref{lemma:anisotropy}) achieved through SymmAligner.

\begin{algorithm}[tp!]
\caption{Consistent Indexing with Dual-Tower Synergy}
\label{alg:indexing}
\begin{algorithmic}[1]
\REQUIRE Trained encoders $f_q$, $f_i$, item collection $\mathcal{D}$, ANN algorithm $\mathcal{A}$
\ENSURE Consistent index for efficient retrieval
\STATE \textbf{Index Construction Phase}:
\FOR{each item $I \in \mathcal{D}$}
    \STATE $\mathbf{e}_I^q \leftarrow f_q(I)$ \COMMENT{Structural vector}
    \STATE $\mathbf{e}_I^i \leftarrow f_i(I)$ \COMMENT{Representation vector}
\ENDFOR
\STATE $\{\mathbf{c}_1, \ldots, \mathbf{c}_K\} \leftarrow \text{k-means}(\{\mathbf{e}_I^q\}, K)$ \COMMENT{Unified clustering}
\FOR{each item $I \in \mathcal{D}$}
    \STATE $k(I) \leftarrow \arg\min_j \|\mathbf{e}_I^q - \mathbf{c}_j\|^2$ \COMMENT{Cluster assignment in unified space}
    \STATE $\mathbf{r}_I \leftarrow \mathbf{e}_I^i - \mathbf{c}_{k(I)}$ \COMMENT{Hybrid-space residual computation}
\ENDFOR
\STATE $\text{index} \leftarrow \mathcal{A}.\text{build}(\{\mathbf{e}_I^q, \mathbf{r}_I: I \in \mathcal{D}\})$ \COMMENT{Algorithm-specific construction}

\STATE \textbf{Retrieval Phase}:
\STATE $\mathbf{e}_Q \leftarrow f_q(Q)$ \COMMENT{Query encoding in unified space}
\STATE $\{I_1, \ldots, I_k\} \leftarrow \mathcal{A}.\text{search}(\mathbf{e}_Q, \text{index}, k)$ \COMMENT{Consistent retrieval}
\RETURN $\{I_1, \ldots, I_k\}$
\end{algorithmic}
\end{algorithm}

\subsubsection{Dual-Vector Strategy and IVF-PQ Implementation}

We implement this framework through a Dual-Vector Strategy that generates specialized embeddings for each item $I \in \mathcal{D}$:
\begin{align*}
\mathbf{e}_I^q &= f_q(I) \quad \text{(Structural Vector)} \\
\mathbf{e}_I^i &= f_i(I) \quad \text{(Representation Vector)}
\end{align*}

In the widely used IVF-PQ scheme, this strategy translates to two key steps:

\begin{enumerate}
    \item \textbf{Unified Clustering}: Cluster centers are computed exclusively from structural vectors to ensure query-space consistency:
    \begin{equation}
    \{\mathbf{c}_1, \ldots, \mathbf{c}_K\} = \text{k-means}(\{\mathbf{e}_I^q\}_{I \in \mathcal{D}}, K)
    \end{equation}
    
    \item \textbf{Hybrid-Space Residual Computation}: Items are assigned to clusters based on structural vectors, but residuals are computed using representation vectors:
    \begin{align}
    k(I) &= \arg\min_j \|\mathbf{e}_I^q - \mathbf{c}_j\|^2  \\
    s.t.  & \quad \mathbf{r}_I= \mathbf{e}_I^i - \mathbf{c}_{k(I)}
    \end{align}
\end{enumerate}

This hybrid approach leverages the strengths of both towers: the structural vectors ensure consistency with query encoding, while the representation vectors potentially capture richer item-specific information for accurate quantization. The mathematical soundness of cross-space residual computation is guaranteed by the encoder equivalence ($f_q(I) \approx f_i(I)$) established through SymmAligner.
\subsubsection{Algorithm-Agnostic Generalization}
The power of the Dual-Vector Strategy lies in its generality. While the IVF-PQ implementation showcases its effectiveness, the principle of separating coarse structure from fine representation is broadly applicable to other major ANN families:
\begin{itemize}
\item \textbf{Quantization methods (IVF-FLAT, IVF-OPQ)}: Use $\mathbf{e}_I^q$ for coarse clustering and the hybrid-space $\mathbf{r}_I$ for product quantization.
\item \textbf{Graph methods (HNSW)}: Build the graph on the structural vectors $\{\mathbf{e}_I^q\}$ to ensure consistent routing, while final distance calculations can optionally use the high-fidelity representation vectors $\mathbf{e}_I^i$ for refinement.
\end{itemize}

\subsubsection{Theoretical Guarantees for Indexing Consistency}

The SymmAligner component provides the theoretical foundation for consistent indexing through representation alignment.

\begin{theorem}[Retrieval Consistency]
\label{theorem:consistency}
Using item representations from the query tower, $\{f_q(I)\}$, for index construction aligns the ANN search objective with the model's true objective.
\end{theorem}

\begin{corollary}[Enhanced Retrieval with Dual-Tower Synergy]
The dual-tower enhanced implementation maintains retrieval consistency while providing improved quantization accuracy through residual computation.
\end{corollary}

These results establish that our indexing framework fundamentally solves the distribution mismatch problem in dual-tower retrieval. Theorem \ref{theorem:consistency} guarantees that approximate nearest neighbor search accurately reflects true semantic similarity, while the corollary demonstrates that our dual-tower enhancement maintains this consistency while improving quantization quality through hybrid-space residual computation.

\subsection{Complexity Analysis}

We analyze the computational complexity of our framework compared to conventional dual-tower models, with key results summarized in Table~\ref{tab:complexity}. The proposed framework introduces carefully balanced trade-offs:
\begin{itemize}
\item Training: Complexity increases from $O(F+B)$ to $O(2F+B)$ due to the additional forward passes for input swapping, though this is partially offset by faster convergence.

\item Index Construction: Offline complexity increases from $O(N \cdot T_i)$ to $O(N \cdot (T_q + T_i))$ as we compute both structural and representation vectors.

\item Online Service: Crucially, all inference metrics remain identical to conventional approaches, preserving real-time performance with $O(D + K \cdot D + n_{\text{probe}} \cdot \frac{N}{K} \cdot M)$ complexity.
\end{itemize}
This design accepts moderate offline costs to achieve significant improvements in representation quality and retrieval consistency, while maintaining deployment efficiency for industrial-scale systems.

\section{Experiments}
In this section, we conduct comprehensive experiments to answer three central research questions (RQs):
\begin{itemize}
\item \textbf{RQ1( Overall Performance):}  Does our SymmAligner and CI framework achieve state-of-the-art performance against baselines?

\item \textbf{RQ2(Component-wise Effectiveness):} What are the individual contributions of SymmAligner and CI, and how do they address the core issues of representation and index inconsistency?

\item \textbf{RQ3(Robustness and Sensitivity):} How robust is our framework to variations in key hyperparameters (e.g., alignment weight $\lambda$)?
\end{itemize}

\subsection{Datasets and Metrics}
Our evaluation is grounded in two settings: a standard public benchmark and a large-scale industrial environment.
\begin{itemize}
    \item \textbf{MS MARCO (MS300k)    \cite{masmarco}:} A widely-used public benchmark for document retrieval
    \cite{d2gen,tsgen,novo,minder}. We follow the preprocessing protocol from \cite{ultron} for URL-based deduplication, using its train/dev sets for our experiments.
    \item \textbf{Industrial E-commerce Dataset:} A proprietary dataset of 10 million user query-item click interactions from a large e-commerce platform, with a corpus of 1 million items. This dataset reflects the challenges of a real-world production environment.
\end{itemize}

For MS MARCO, we report standard metrics \cite{tsgen,d2gen,DSI,cheng2025descriptive}: Mean Reciprocal Rank (MRR@K) and Recall@K. For the industrial dataset, we focus on metrics crucial for large-scale candidate generation: Precision@K, Recall@K, and NDCG@K.

\subsection{Baselines}
Our comparative analysis includes three categories of retrieval methods.
\begin{itemize}
    \item \textbf{Sparse retrieval:} BM25, and advanced learned sparse models like UniCOIL \cite{unicoil} and SPLADEv2 \cite{SPLADEv2}.
    \item \textbf{Standard dense retrieval:} DPR \cite{karpukhin2020dense}, ANCE \cite{ANCE}, and critically, our own \textbf{Bert2Tower}, which uses the BERT backbone and training setup as our method (sans SymmAligner and CI) to ensure a fair and direct comparison.
    \item \textbf{Enhanced dense retrieval:} GTR-Base \cite{gpt-base}, a pre-trained generalizable T5-based dense retrievers.
\end{itemize}

Our work focuses on enhancing the standard, pre-computation dual-tower paradigm, which is fundamental to industrial-scale, low-latency retrieval systems. We therefore do not directly compare against methods that alter this paradigm. Late-interaction models (e.g., ColBERT \cite{khattab2020colbert}) sacrifice pre-computation efficiency for higher-cost runtime interactions. Knowledge distillation methods rely on computationally expensive teacher models (e.g., Cross-Encoders), whereas our approach is self-contained. Joint optimization methods typically focus on minimizing quantization loss but do not address the precedent, more fundamental problem of representation space misalignment that our framework targets.

Note that, we did not employ a parameter-sharing architecture due to the inherent data heterogeneity (e.g., text queries vs. multimodal items) and architectural asymmetry (a lightweight online query tower vs. a complex offline item tower). This dual conflict induces a "seesaw effect," which degrades the representation quality for both sides. We confirmed this empirically on an industrial dataset, where our parameter-sharing model, Bert2Tower-Shared, showed a drop in performance.

\subsection{Implementation Details}
All dense retrievers are built upon the BERT encoder architecture \cite{devlin2019bert}. We use the AdamW optimizer with a constant learning rate $1 \times 10^{-5}$, preceded by a linear warmup phase. The embedding dimension is 768. For SymmAligner, the alignment weight $\lambda$ was set to 0.3 based on a grid search on a validation set. For ANN indexing, we use IVF-PQ\cite{johnson2019billion} with $nlist=4096$, $nprobe=64$, and $M=64$ (64-byte codes), unless stated otherwise.

\begin{table*}[htbp!]
\centering
\caption{Main results on the MS MARCO (MS300K) dataset. \textbf{Bert2Tower} is our baseline dual-tower model. \textbf{SCI (Bert2Tower+ SymmAligner)} is the upper bound of our SymmAligner-trained model with exact search. \textbf{Proposed ($\lambda$=...)} are our full practical frameworks (SCI) with different alignment weights.}
\resizebox{0.9\textwidth}{!}{
\begin{tabular}{llccccc}
\toprule
                            &                       & \multicolumn{5}{c}{MS MARCO (MS300K)}                                \\ 
\cmidrule(lr){3-7} 
Category                    & Method                & MRR@10  & MRR@100 & Recall@1   & Recall@10  & Recall@100 \\ 
\midrule
\multirow{3}{*}{Sparse}     & BM25                  & 0.248 & 0.255   & 0.186 & 0.391 & 0.573 \\
                            & UniCOIL               & 0.425 & 0.435   & 0.284 & 0.766 & 0.951 \\
                            & SPLADEv2              & 0.443 & 0.452   & 0.328 & 0.779 & 0.956 \\ 
\midrule
\multirow{5}{*}{Dense(Brute-Force)}      & DPR                   & 0.424 & 0.433   & 0.271 & 0.764 & 0.948 \\
                            & ANCE                  & 0.451 & 0.455   & 0.299 & 0.785 & 0.953 \\
                            & GTR-Base              & 0.576 & 0.581   & 0.471 & 0.785 & 0.912 \\ 
                        
\cmidrule(lr){2-7} 
                            & Bert2Tower         & 0.480 & 0.488   & 0.372 & 0.714 & 0.893 \\ 
                      
                             & \textbf{SCI (Bert2Tower+ SymmAligner) }       & 0.496 & 0.504   & 0.383 & 0.737 & 0.908 \\ 
\midrule
\multirow{6}{*}{Dense(IVF-Flat)}     & Bert2Tower       & 0.408 & 0.414   & 0.318 & 0.603 & 0.750 \\
\cmidrule(lr){2-7} 

                            & SCI ($\lambda=0.1$) & 0.369 & 0.374   & 0.287 & 0.545 & 0.674 \\
                            & SCI ($\lambda=0.2$) & 0.400 & 0.407   & 0.316 & 0.580 & 0.721 \\
                            & \textbf{SCI ($\lambda=0.3$}) & \textbf{0.448} & \textbf{0.455} & \textbf{0.350} & \textbf{0.657} & \textbf{0.805} \\
                            & SCI ($\lambda=0.4$) & 0.420 & 0.426   & 0.324 & 0.620 & 0.756 \\
                            & SCI ($\lambda=0.5$) & 0.393 & 0.399   & 0.303 & 0.588 & 0.717 \\
\bottomrule
\end{tabular}
}
\label{pub_exp}
\end{table*}

\subsection{Overall Results (RQ1 and RQ3)}

Table \ref{pub_exp} presents the main results of our framework against various baselines on the MS MARCO dataset, including a detailed ablation study on the SymmAligner alignment weight, $\lambda$. The results validate our framework's effectiveness through three key findings.

First, our full, practical framework demonstrates a significant performance advantage over the comparable baseline. Our best model, \textbf{SCI ($\lambda=0.3$)}, achieves an MRR@10 of \textbf{0.448}, which is a substantial \textbf{9.9\% relative improvement} over the \textbf{Bert2Tower} baseline (0.408). This direct, apples-to-apples comparison on indexed systems confirms that our end-to-end consistency-oriented approach (SCI) effectively resolves the bottlenecks in standard dual-tower retrieval, leading to superior ranking quality.

Second, the results clearly delineate the two-stage contribution of our framework. The brute-force evaluation of \textbf{SCI (Only SymmAligner)} (MRR@10=0.496) shows a marked improvement over the \textbf{Bert2Tower} counterpart (0.480), confirming that SymmAligner successfully creates a higher-quality representation space. While ANN indexing inevitably introduces a performance gap, our final model (MRR@10=0.448) successfully preserves a large portion of the gains established by SymmAligner, demonstrating the crucial role of CI in translating theoretical representation quality into practical performance.

Finally, the ablation on the alignment weight $\lambda$ reveals a non-trivial trade-off and offers important insights. Performance steadily improves as $\lambda$ increases from 0.1 to 0.3, peaking at our best model. However, as $\lambda$ increases further to 0.5, performance begins to decline. This demonstrates that while the symmetric alignment signal from $\mathcal{L}_{\text{swap}}$ is critical, an excessive alignment constraint can start to interfere with the primary retrieval objective of $\mathcal{L}_{\text{original}}$. The existence of an optimal "sweet spot" at $\lambda=0.3$ empirically validates our approach of balancing these two loss components, providing a clear and practical methodology for optimizing consistency in dual-tower models.

\begin{table}[tbp!]
    \centering
    \caption{
        Cosine similarity for ground-truth query-item pairs on the MS MARCO test set.
    }
    \label{tab:gt_similarity}
    \resizebox{0.4\textwidth}{!}{
    \begin{tabular}{lcc}
        \toprule
        Statistic Metric & Bert2Tower& \textbf{SymmAligner} \\
        \midrule
        Mean Similarity    & 0.615 & \textbf{0.672} \\
        Median Similarity  & 0.625 & \textbf{0.684} \\
        \midrule
        Minimum Similarity & 0.232 & \textbf{0.364} \\
        Maximum Similarity & 0.783 & \textbf{0.830} \\
        \midrule
        Std. Deviation     & 0.066 & \textbf{0.065} \\
        \bottomrule
    \end{tabular}
    }
    \vspace{-2mm}
\end{table}

\subsection{Analysis on SymmAligner (RQ2)}
As shown in Table \ref{tab:gt_similarity}, the baseline model achieves a mean similarity of 0.615 for ground-truth pairs. Our SymmAligner-enhanced model increases this value to \textbf{0.672}, a relative improvement of 9.3\%. This quantitative increase in average similarity provides direct empirical validation for \textbf{Lemma \ref{lemma:alignment}}, demonstrating a more effective and consistent alignment of positive pairs.

Of particular significance is the change in the minimum similarity, which increases by over 50\% from 0.232 to \textbf{0.364}. The baseline's lower value indicates the presence of challenging positive pairs that are mapped to geometrically distant regions in the embedding space. The substantial improvement in this worst-case alignment suggests a more robust and uniform semantic manifold. By significantly improving the scores of these hardest pairs, SymmAligner mitigates the detrimental effects of representation inconsistencies—a key goal related to \textbf{Lemma \ref{lemma:anisotropy}}—and reduces the risk of catastrophic ranking failures. 

\begin{table*}[htbp!]
\centering
\caption{Performance comparison with different nprobe values.}
\resizebox{0.9\textwidth}{!}{
\begin{tabular}{ll|ccc|ccc|ccc|ccc}
\toprule
\multicolumn{2}{c|}{\multirow{2}{*}{Method}} & \multicolumn{3}{c|}{Precision} & \multicolumn{3}{c|}{Recall} & \multicolumn{3}{c|}{MRR} & \multicolumn{3}{c}{NDCG} \\
\cmidrule(lr){3-5} \cmidrule(lr){6-8} \cmidrule(lr){9-11} \cmidrule(lr){12-14}
 & & @1 & @10 & @100 & @1 & @10 & @100 & @1 & @10 & @100 & @1 & @10 & @100 \\
\midrule
\multirow{2}{*}{nprobe=1} & Bert2Tower & 0.1348 & 0.0277 & 0.0035 & 0.1348 & 0.2767 & 0.3476 & 0.1348 & 0.1771 & 0.1807 & 0.1348 & 0.0528 & 0.0571 \\
 & SCI($\lambda=0.3$) & 0.1662 & 0.0327 & 0.0039 & 0.1662 & 0.3268 & 0.3929 & 0.1662 & 0.2157 & 0.2189 & 0.1662 & 0.0637 & 0.0677 \\
\midrule
\multirow{2}{*}{nprobe=8} & Bert2Tower & 0.2634 & 0.0502 & 0.0062 & 0.2634 & 0.5020 & 0.6241 & 0.2634 & 0.3372 & 0.3429 & 0.2634 & 0.0988 & 0.1060 \\
 & SCI($\lambda=0.3$) & 0.2770 & 0.0530 & 0.0065 & 0.2770 & 0.5300 & 0.6478 & 0.2770 & 0.3560 & 0.3613 & 0.2770 & 0.1043 & 0.1112 \\
\midrule
\multirow{2}{*}{nprobe=16} & Bert2Tower & 0.2952 & 0.0556 & 0.0069 & 0.2952 & 0.5564 & 0.6933 & 0.2952 & 0.3766 & 0.3829 & 0.2952 & 0.1101 & 0.1180 \\
 & SCI($\lambda=0.3$) & 0.3054 & 0.0578 & 0.0071 & 0.3054 & 0.5782 & 0.7083 & 0.3054 & 0.3921 & 0.3980 & 0.3054 & 0.1145 & 0.1220 \\
\midrule
\multirow{2}{*}{nprobe=32} & Bert2Tower & 0.3181 & 0.0603 & 0.0075 & 0.3181 & 0.6030 & 0.7501 & 0.3181 & 0.4077 & 0.4144 & 0.3181 & 0.1192 & 0.1277 \\
 & SCI($\lambda=0.3$) & 0.3291 & 0.0617 & 0.0076 & 0.3291 & 0.6171 & 0.7575 & 0.3291 & 0.4202 & 0.4265 & 0.3291 & 0.1225 & 0.1306 \\
\midrule
\multirow{2}{*}{nprobe=64} & Bert2Tower & 0.3403 & 0.0641 & 0.0080 & 0.3403 & 0.6410 & 0.7968 & 0.3403 & 0.4353 & 0.4423 & 0.3403 & 0.1270 & 0.1360 \\
 & SCI($\lambda=0.3$) & 0.3501 & 0.0657 & 0.0081 & 0.3501 & 0.6568 & 0.8053 & 0.3501 & 0.4480 & 0.4547 & 0.3501 & 0.1305 & 0.1391 \\
\bottomrule
\end{tabular}
}
\label{tab:nprobe_comparison}
\end{table*}

\begin{table*}[htbp!]
\centering
\caption{Performance on the industrial dataset. 
    The top section shows brute-force search results. The bottom section shows results with an IVF-PQ index.
}
\resizebox{0.9\textwidth}{!}{
\begin{tabular}{l|l|ccc|ccc|ccc|ccc}
\toprule
\multirow{2}{*}{Category} & \multirow{2}{*}{Method} & \multicolumn{3}{c|}{Precision} & \multicolumn{3}{c|}{Recall} & \multicolumn{3}{c|}{MRR} & \multicolumn{3}{c}{NDCG} \\
\cmidrule(lr){3-5} \cmidrule(lr){6-8} \cmidrule(lr){9-11} \cmidrule(lr){12-14}
 & & @10 & @100 & @1000 & @10 & @100 & @1000 & @10 & @100 & @1000 & @10 & @100 & @1000 \\
\midrule
\multirow{3}{*}{Dense (Brute-Force)} & Bert2Tower-Shared & 0.0579 & 0.0114 & 0.0014 & 0.4286 & 0.7997 & 0.9664 & 0.2268 & 0.2412 & 0.2420 & 0.2657 & 0.3510 & 0.3746 \\
                   & Bert2Tower & 0.0588 & 0.0115 & 0.0014 & 0.4344 & 0.8056 & 0.9683 & 0.2367 & 0.2514 & 0.2521 & 0.2739 & 0.3594 & 0.3823 \\
                   & SCI w/o CI & \textbf{0.0594} & \textbf{0.0115} & \textbf{0.0014} & \textbf{0.4389} & \textbf{0.8065} & \textbf{0.9678} & \textbf{0.2373} & \textbf{0.2516} & \textbf{0.2523} & \textbf{0.2755} & \textbf{0.3600} & \textbf{0.3828} \\
\midrule
\multirow{4}{*}{Dense (IVF-PQ)} & Bert2Tower & 0.0322 & 0.0085 & 0.0013 & 0.2378 & 0.5984 & 0.8881 & 0.1182 & 0.1317 & 0.1329 & 0.1385 & 0.2181 & 0.2582 \\
                   & SCI ($\lambda=0.1$) & 0.0322 & 0.0083 & 0.0012 & 0.2392 & 0.5814 & 0.8627 & 0.1234 & 0.1362 & 0.1374 & 0.1430 & 0.2188 & 0.2575 \\
                   & \textbf{SCI ($\lambda=0.3$)} & \textbf{0.0368} & \textbf{0.0089} & 0.0013 & \textbf{0.2718} & \textbf{0.6222} & 0.8793 & \textbf{0.1360} & \textbf{0.1493} & \textbf{0.1503} & \textbf{0.1603} & \textbf{0.2384} & \textbf{0.2740} \\
                   & SCI ($\lambda=0.5$) & 0.0229 & 0.0073 & 0.0012 & 0.1706 & 0.5162 & 0.8571 & 0.0798 & 0.0922 & 0.0937 & 0.0962 & 0.1705 & 0.2177 \\
\bottomrule
\end{tabular}
}
\label{tab:performance_comparison}
\end{table*}

\subsection{Analysis on CI (RQ2)}
We validate CI by analyzing its performance under varying search scopes controlled by \texttt{nprobe} (Table \ref{tab:nprobe_comparison}). The \texttt{nprobe} parameter, representing the computational cost of the coarse search, allows for a direct evaluation of the index's accuracy and efficiency.

As the results show, at a minimal search scope of \texttt{nprobe=1}, our method improves Recall@10 by 18\% relative to the traditional index (\texttt{Bert2Tower}). This demonstrates a more accurate coarse-grained lookup, as the initial cluster searched in the CI index is more likely to contain relevant items. This improved accuracy translates to greater overall search efficiency. For example, our CI at \texttt{nprobe=8} achieves an MRR@100 of 0.3613, a value comparable to the baseline which requires a larger search scope of \texttt{nprobe=16} to reach a similar score of 0.3829. This indicates that CI can achieve a similar level of ranking quality with approximately half the computational cost. The ability to attain higher accuracy with a smaller search scope empirically validates \textbf{Theorem \ref{theorem:consistency}}, confirming that our consistent indexing creates a more effective and efficient retrieval path.

\subsection{Results on Industrial-Scale Data}

To validate our framework's practical value and scalability, we present results from a large-scale industrial e-commerce dataset in Table \ref{tab:performance_comparison}. The baseline results starkly illustrate the core challenge addressed by this paper: a massive performance gap exists between the model's brute-force potential (\texttt{Bert2Tower}, R@100=0.8056) and its practical performance after applying a standard IVF-PQ index (\texttt{Bert2Tower}, R@100=0.5984). This gap underscores the severe performance degradation caused by representation and indexing inconsistencies in a real-world production system.

Our consistency-oriented framework is designed to bridge this gap. Our full system, \texttt{SCI ($\lambda=0.3$)}, significantly improves upon the indexed baseline, achieving a \textbf{4.0\% relative gain in Recall@100} and, more critically, a \textbf{9.3\% relative gain in NDCG@100}. The substantial NDCG lift provides a deeper insight that corroborates our findings on the public benchmark: our framework not only retrieves more relevant items but also ranks them more accurately at the top. This is the direct empirical outcome of our systematic, two-part solution. SymmAligner first creates a more robust representation space (evidenced by the consistent gains of \texttt{SCI w/o CI} in the brute-force evaluation), and CI then ensures this improved semantic structure is preserved during the coarse-to-fine ANN search, as validated by our \texttt{nprobe} analysis. The results also reveal an optimal alignment strength ($\lambda=0.3$), offering practical guidance. Ultimately, the industrial data provides strong evidence that our end-to-end consistency framework is highly effective at translating a model's theoretical potential into tangible real-world performance.

\section{Conclusion}
In this paper, we mitigate the problem of representational and indexing inconsistency in dual-tower dense retrieval systems, a critical bottleneck for both traditional and emerging generative recommendation paradigms. We introduce a novel consistency-oriented framework featuring two synergistic components: Symmetric Representation Alignment (SymmAligner) to forge a unified, isotropic semantic space, and Consistent Indexing with dual-tower synergy (CI) to construct a consistent retrieval index. Extensive experiments on public benchmarks and a billion-scale industrial dataset validate that our integrated approach significantly improves retrieval performance by systematically eliminating the mismatch between training and inference.

\bibliographystyle{ACM-Reference-Format}
\bibliography{sample-base}

\appendix
 \fancyhead[RO]{} 
\fancyhead[LE]{} 
\fancyhead[LO, RE]{}  

\section{Complexity Table}
\begin{table*}[htbp]
\centering
\caption{Complexity Comparison Analysis}
\label{tab:complexity}
\begin{tabular}{lcc}
\toprule
\textbf{Dimension} & \textbf{Conventional Two-Tower} & \textbf{Proposed Method} \\
\midrule
Index and Retrieval & $\{I_1, \ldots, I_k\} = \mathcal{A}\left(f_q(Q), \{f_i(I)\}_{I \in \mathcal{D}}\right)$  & $\{I_1, \ldots, I_k\} = \mathcal{A}\left(f_q(Q); \underbrace{\{f_q(I)\}_{I \in \mathcal{D}}}_{\text{Coarse Structure}} , \underbrace{\{f_i(I)\}_{I \in \mathcal{D}}}_{\text{Fine Representation}}\right)$  \\
Training Complexity & $O(F+B)$ & $O(2F+B)$ \\
Index Construction & $O(N \cdot T_i)$ & $O(N \cdot (T_q + T_i))$  \\
Index Storage & $O(KD + MK^*D^* + NM)$ & Identical \\
Online Inference & $O(D + K \cdot D + n_{\text{probe}} \cdot \frac{N}{K} \cdot M)$ & Identical\\
Memory Consumption & $O(P + M_{\text{Index}})$ & Identical \\
\bottomrule
\end{tabular}
\end{table*}

\section{Proofs of Theoretical Results}
\label{app:proofs}

\subsection{Proof of Lemma \ref{lemma:independence}}
\label{app:lemma1}

\begin{proof}
Let $J_q(X) = \nabla_{\theta_q}f_q(X)$ be the Jacobian of the query tower. The gradients are expectations over Jacobians evaluated at different data distributions:
\begin{align*}
\nabla_{\theta_q}\mathcal{L}_{\text{original}} &\propto \mathbb{E}_{Q,I}\left[J_q(Q)^\top \cdot g_1(f_q(Q), f_i(I))\right] \\
\nabla_{\theta_q}\mathcal{L}_{\text{swap}} &\propto \mathbb{E}_{Q,I}\left[J_q(I)^\top \cdot g_2(f_i(Q), f_q(I))\right]
\end{align*}
where $g_1$ and $g_2$ are different gradient functions derived from the loss formulations.

Since the expectations are taken over Jacobians corresponding to inputs from different distributions ($Q$ vs. $I$), and assuming query and item distributions are not identical, the resulting gradient vectors are not scalar multiples of each other. This ensures $\mathcal{L}_{\text{swap}}$ provides complementary, non-redundant optimization signals.

More formally, if we assume $\nabla_{\theta_q}\mathcal{L}_{\text{swap}} = \alpha \nabla_{\theta_q}\mathcal{L}_{\text{original}}$ for some scalar $\alpha$, this would imply that the Jacobians evaluated at query and item inputs are linearly dependent, which contradicts the assumption that query and item distributions are distinct and the encoders are sufficiently expressive.
\end{proof}

\subsection{Proof of Lemma \ref{lemma:alignment}}
\label{app:lemma2}

\begin{proof}
At an optimal parameterization $\theta^*$, both loss components are minimized, i.e., $\mathcal{L}_{\text{original}}(\theta^*) \to 0$ and $\mathcal{L}_{\text{swap}}(\theta^*) \to 0$. This implies that for any given triplet, the following conditions must hold:
\begin{align}
    S(f_q(Q), f_i(I^+)) - S(f_q(Q), f_i(I^-)) &\ge \delta \label{eq:cond1} \\
    S(f_i(Q), f_q(I^+)) - S(f_i(Q), f_q(I^-)) &\ge \delta \label{eq:cond2}
\end{align}

Let $\Delta_{\text{orig}}$ and $\Delta_{\text{swap}}$ be the score differences in (\ref{eq:cond1}) and (\ref{eq:cond2}) respectively. The optimization encourages both $\Delta_{\text{orig}}$ and $\Delta_{\text{swap}}$ to be close to the margin $\delta$. Consequently, their difference is driven towards zero in expectation:
\begin{equation}
\mathbb{E}[\Delta_{\text{orig}} - \Delta_{\text{swap}}] \to 0
\end{equation}

Expanding this term:
\begin{align*}
\Delta_{\text{orig}} - \Delta_{\text{swap}} = &\left[S(f_q(Q), f_i(I^+)) - S(f_q(Q), f_i(I^-))\right] \\
& - \left[S(f_i(Q), f_q(I^+)) - S(f_i(Q), f_q(I^-))\right]
\end{align*}

For this difference to approach zero over the entire data distribution, the individual similarity functions must converge, i.e.,:
\begin{equation}
S(f_q(Q), f_i(I)) \approx S(f_i(Q), f_q(I)) \quad \forall Q,I
\end{equation}

This directly implies the minimization of the alignment error $\mathcal{A}(\theta_q, \theta_i)$ as defined in the lemma.
\end{proof}

\subsection{Proof of Lemma \ref{lemma:anisotropy}}
\label{app:lemma3}

\begin{proof}
Anisotropy can be quantified by the condition number of the embedding covariance matrix $\Sigma = \text{Cov}(f(X))$, i.e., $\mathcal{H} = \lambda_{\max}(\Sigma) / \lambda_{\min}(\Sigma)$. A high $\mathcal{H}$ implies embeddings occupy a low-dimensional subspace.

Without $\mathcal{L}_{\text{swap}}$, the model can minimize $\mathcal{L}_{\text{original}}$ by creating two separate anisotropic cones for queries and items, i.e., finding $\Sigma_q$ and $\Sigma_i$ that are individually ill-conditioned.

The inclusion of $\mathcal{L}_{\text{swap}}$ forces a shared semantic structure. This imposes a consistency constraint on the covariance matrices, as the query encoder must handle the item distribution and vice versa. The optimization is incentivized to find a solution where the covariance structures are compatible:
\begin{equation}
\text{Cov}(f_q(\mathcal{Q} \cup \mathcal{I})) \approx \text{Cov}(f_i(\mathcal{Q} \cup \mathcal{I}))
\end{equation}

This constraint makes it difficult to maintain separate, ill-conditioned subspaces. Instead, the model is pushed to utilize the embedding dimensions more uniformly, resulting in a lower condition number $\mathcal{H}$ and a more isotropic space.

More formally, consider the combined covariance matrix:
\begin{equation}
\Sigma_{\text{combined}} = \frac{1}{2}\left(\text{Cov}(f_q(\mathcal{Q} \cup \mathcal{I})) + \text{Cov}(f_i(\mathcal{Q} \cup \mathcal{I}))\right)
\end{equation}

The swap loss encourages the individual covariance matrices to be close to this combined matrix, preventing extreme anisotropy in either tower's representation space.
\end{proof}

\subsection{Proof of Theorem \ref{theorem:consistency}}
\label{app:theorem1}

\begin{proof}
Let the ideal search be $i^* = \arg\max_I S(f_q(Q), f_i(I))$. The ANN search approximates this by first selecting a candidate set $\mathcal{C}_Q = \{I \mid \mathcal{P}(f_q(I)) \in \mathcal{K}_Q\}$, where $\mathcal{P}$ is the cluster assignment function and $\mathcal{K}_Q$ is the set of probed clusters for query $Q$. Inconsistency arises when $i^* \notin \mathcal{C}_Q$.

Our proposed training and indexing method minimizes the probability of this event through three key mechanisms:

\begin{enumerate}
    \item From the encoder equivalence induced by representation alignment (Lemma \ref{lemma:alignment}), we have $f_i(I) \approx f_q(I)$. The search objective is therefore well-approximated by an intra-tower problem:
    \begin{equation}
    \arg\max_I S(f_q(Q), f_i(I)) \approx \arg\max_I S(f_q(Q), f_q(I))
    \end{equation}
    
    \item From the anisotropy reduction (Lemma \ref{lemma:anisotropy}), the embedding space of $f_q$ is more isotropic. In such a space, dot product similarity is strongly correlated with Euclidean distance for normalized vectors:
    \begin{equation}
    \arg\max_I S(f_q(Q), f_q(I)) \approx \arg\min_I \|f_q(Q) - f_q(I)\|_2
    \end{equation}
    
    \item The ANN coarse search, which selects clusters based on $\min_{IC} \|f_q(Q) - IC\|_2$, where $IC$ are centroids of $\{f_q(I)\}$, is a quantized approximation of the RHS of the equation above.
\end{enumerate}

\fancyhead[RO]{} 
\fancyhead[LE]{} 
\fancyhead[LO, RE]{}  

Thus, the coarse search objective becomes a valid proxy for the true ranking objective, ensuring retrieval consistency. The approximation error decreases as the number of clusters increases and the representation alignment improves.
\end{proof}

\subsection{Proof of Corollary 1}
\label{app:corollary1}

\begin{proof}
The dual-tower enhanced implementation maintains consistency through:

\begin{enumerate}
    \item \textbf{Coarse clustering consistency}: The structural vectors $\{\mathbf{e}_I^q\}$ are used for cluster assignment, ensuring alignment with query encoding as established in Theorem \ref{theorem:consistency}.
    
    \item \textbf{Residual semantic preservation}: The hybrid-space residual computation $\mathbf{r}_I = \mathbf{e}_I^i - \mathbf{c}_{k(I)}$ preserves semantic relationships due to $\mathbf{e}_I^i \approx \mathbf{e}_I^q$ from representation alignment (Lemma \ref{lemma:alignment}). This ensures that:
    \begin{equation}
    \|\mathbf{e}_Q^q - \mathbf{e}_I^i\| \approx \|\mathbf{e}_Q^q - \mathbf{e}_I^q\| + \|\mathbf{r}_I\|
    \end{equation}
    
    \item \textbf{Enhanced quantization}: The representation vectors $\{\mathbf{e}_I^i\}$ potentially capture richer item-specific information, leading to more accurate residual quantization while maintaining the unified space constraint through the hybrid-space formulation.
\end{enumerate}

The combination of these factors ensures that retrieval consistency is maintained while quantization accuracy is improved through the dual-tower synergy.
\end{proof}

\subsection{Detailed Collapse Analysis for Linear Model}
\label{app:collapse}

The total gradient for the linear model is:
\begin{equation}
\frac{\partial \mathcal{L}_{\text{total}}}{\partial W_q} = W_i \left[ \Delta i q^\top + \lambda q \Delta i^\top \right]
\end{equation}
where $\Delta i = i^- - i^+$.

We analyze the conditions under which the swap loss contribution becomes redundant:

\subsubsection{Collapse Condition 1: $\lambda = 0$}
When $\lambda = 0$, the swap loss has no weight, and the total gradient degenerates to:
\begin{equation}
\frac{\partial \mathcal{L}_{\text{total}}}{\partial W_q} = W_i \Delta i q^\top
\end{equation}
This is identical to the original gradient, and the swap loss produces no optimization effect.

\subsubsection{Collapse Condition 2: Vector Parallelism ($q \parallel \Delta i$)}
When the query vector $q$ is parallel to the document difference vector $\Delta i$ (i.e., $q = \alpha \Delta i$ for some scalar $\alpha$), the symmetrization operator degenerates:
\begin{equation}
M(q, \Delta i) = \frac{1}{2}(q \Delta i^\top + \Delta i q^\top) = \frac{\alpha + 1}{2} \Delta i \Delta i^\top
\end{equation}
The total gradient becomes:
\begin{equation}
\frac{\partial \mathcal{L}_{\text{total}}}{\partial W_q} = W_i \left[ \Delta i q^\top + \lambda q \Delta i^\top \right] = W_i (\alpha + \lambda \alpha^2) \Delta i \Delta i^\top
\end{equation}
This is proportional to the original gradient, making the swap gradient redundant.

\subsubsection{Collapse Condition 3: Special Orthogonal Conditions}
When $W_i q$ and $W_i \Delta i$ are orthogonal in the transformed space, the two gradient directions may produce cancellation effects in certain dimensions. Specifically, if:
\begin{equation}
(W_i q)^\top (W_i \Delta i) = 0
\end{equation}
then the outer products $\Delta i q^\top$ and $q \Delta i^\top$ have orthogonal row spaces in the transformed domain, potentially leading to dimensional cancellation.

\subsubsection{Practical Implications}
In practice, these collapse conditions are rarely fully satisfied:
\begin{itemize}
    \item $\lambda$ is typically set in $[0.1, 1.0]$, avoiding complete removal of swap loss
    \item Queries and items typically come from different semantic spaces with different statistical properties, making $q$ and $\Delta i$ rarely completely parallel
    \item The orthogonal condition requires specific geometric relationships that are unlikely to hold across the entire dataset
\end{itemize}

Therefore, the swap loss generally provides meaningful complementary optimization signals that promote representation space alignment in practical scenarios.

\end{document}